%% file: draft.tex
\documentclass[aps,prr,twocolumn,showpacs,floatfix,superscriptaddress,amsmath,amssymb]{revtex4-2}
\usepackage{graphicx}
\usepackage[dvipsnames]{xcolor}
\usepackage{bm}
\usepackage{enumitem}
\usepackage{microtype}
\usepackage[pdfusetitle]{hyperref}
\usepackage[capitalize]{cleveref}

\crefname{equation}{Eq.}{}
\crefrangeformat{equation}{Eqs.~(#3#1#4)-(#5#2#6)}

\graphicspath{{figs}}

\usepackage[disable]{todonotes}
\setuptodonotes{inline}

\input{abbreviations.tex}

\newcommand{\includegraphicsprint}[2][]{\fbox{\url{#2}}}

\begin{document}

\title{Disorder to Order Transition in 1D non-reciprocal Cahn-Hilliard Model}
\author{Navdeep Rana}
\email{navdeep.rana@ds.mpg.de}
\affiliation{Max Planck Institute for Dynamics and Self-Organization (MPI-DS), D-37077 G\"ottingen, Germany}

\author{Ramin Golestanian}
\email{ramin.golestanian@ds.mpg.de}
\affiliation{Max Planck Institute for Dynamics and Self-Organization (MPI-DS), D-37077 G\"ottingen, Germany}
\affiliation{Rudolf Peierls Centre for Theoretical Physics, University of Oxford, Oxford OX1 3PU, United Kingdom}

\begin{abstract}
    We present the phenomenology of the one dimensional non-reciprocal Cahn Hilliard model for varying non-reciprocity $(\alpha)$ and different boundary conditions. At small $\alpha$, a perturbed uniform state evolves to a defect laden configuration that lacks global polar order. Defects are the sources and sinks of travelling waves. For a given $\alpha$, defects with a unique wave number that increases monotonically with $\alpha$ are selected. A critical threshold $\alphac$ marks the onset of a transition to states with finite global polar order. For periodic boundary conditions, above $\alphac$, the system shows travelling waves that are completely ordered. In contrast, travelling waves are incompatible with the Neumann and Dirichlet boundary conditions. Instead, for $\alpha \gtrsim \alphac$, we find fluctuating domains that show intermittent polar order and at large $\alpha$, the system partitions into two domains with opposite polar order.
\end{abstract}

\maketitle
\listoftodos

\section{Introduction\label{sec:introduction}}
The non-reciprocal Cahn-Hilliard (NRCH) model  describes phase separation of multi-component mixtures in the presence of non-reciprocal couplings~\cite{saha2020, you2020, rana2024a, rana2024b, parkavousi2025, pisegna2024, pisegna2025, saha2025}. Systems with non-reciprocity are intrinsically out of equilibrium and show rich phenomenology~\cite{loos2020, osat2023, osat2024, soto2014, ivlev2015, agudo-canalejo2019, fruchart2021, golestanian2024}. In the NRCH model, the spontaneous breaking of parity and time-reversal symmetries leads to the formation of travelling density bands~\cite{saha2020, you2020}, suppressed coarsening dynamics~\cite{frohoff-hulsmann2021, saha2020}, localized states~\cite{frohoff-hulsmann2021a}, chaotic steady states~\cite{saha2025}, true long-range polar order in two dimensions~\cite{pisegna2025}, and enhanced stability in multispecies mixtures~\cite{parkavousi2025}. Introduced phenomenologically, the NRCH model emerges as a universal amplitude equation for a conserved-Hopf instability in systems with two conservation laws~\cite{frohoff-hulsmann2023} and can also be derived from a systematic coarse-graining of a microscopic model of phoretically active Janus colloids~\cite{golestanian2022, tucci2024}. Non-reciprocal interactions emerge in a variety of non-equilibrium systems with hydrodynamic interactions \cite{Uchida2010_PRL} and vision-cone interactions \cite{LavergneBechinger2019,cohen2014,Khadka2018}, active mixtures \cite{soto2014,soto2015,sahaNJP_2019,agudo-canalejo2019,LokrshiMaitraPRE,fruchart2021,Ouazan-Reboul2023}, mixtures of active and passive components~\cite{WittkowskiNJP2017,Tayar2023}, and some mass-conserved reaction-diffusion systems \cite{Min2Densities_Brauns2021,Wurthner2022}. 

Studies of the NRCH model in two dimensions (2D) have revealed a rich phenomenology due to the presence of defects, namely, spirals with unit magnitude topological charge and topologically neutral targets~\cite{rana2024a, rana2024b}.
These defects are the sources of the travelling waves, and thus the precursors of global polar order.
For a given strength of non-reciprocal coupling $\alpha$, defects with a unique asymptotic wave number, $\kl \propto \sqrt{\alpha}$ are selected~\cite{rana2024a}.
At a critical threshold of nonreciprocity $\alphac$; we find a disorder-to-order transition.
For $\alpha < \alphac$, random disordered states evolve to quasi-stationary defect networks with no global polar order. On the other hand, for $\alpha > \alphac$ the system admit travelling waves without any defects.
Studying the interactions between pairs of defects reveals that the stability of targets in this model fundamentally alters the nature of dynamics of the defects~\cite{rana2024b}, as compared with previously studied nonequilibrium systems with nonconserved order parameter~\cite{aranson1993, aranson2002, hagan1981, hagan1982}.
Thus, the NRCH model provides a fertile ground to investigate the defect dynamics in non-reciprocal systems with conserved order parameter.

In this paper, we present an extensive study on the defects of the NRCH model in one dimension (1D).
As compared to our previous investigations in 2D~\cite{rana2024a, rana2024b}, which have showcased the complexity of defect dynamics in the NRCH model, the 1D configuration provides the opportunity to investigate the role of the defects in a simpler setting, for a number of reasons: (i) Topologically charged defects, which would be the analogue of spirals in 2D, are ruled out. (ii) Sources of travelling waves in 1D are the analogue of targets in 2D. Similarly, sinks are the 1D analogue of domain walls. 
(iii) Waves emanated from the sources end at the sinks; thus sources and sinks are arranged in an alternating manner on the 1D domain. Furthermore, the defect motion is also limited to the 1D line. (iv) Since 1D allows only for sources and sinks, defect interactions are also simplified. At small separations, a source-and-sink pair annihilates. At larger distances, they form stable configurations. These reasons afford us a deeper and more pedagogical representation of the interplay between defect dynamics and the emergent properties of the NRCH model. 

In addition to these effects, the 1D geometry allows us to explore the role of boundary conditions in the dynamics of the NRCH model. While the disorder-to-order transition at higher $\alpha$ leads to travelling waves for periodic boundary conditions, pure unidirectional travelling waves are incompatible with Neumann and Dirichlet boundary conditions, which are more relevant to experimental systems. How the transition proceeds in such cases and what kind of phases emerge from these transitions remain unknown. We investigate the defect dynamics and the effect of the boundary conditions on the disorder-order transition in the 1D NRCH model under different boundary conditions. Our main results are as follows: (i) We find a transition from disordered, defect laden states to polar ordered states. (ii) Below the transition threshold ($\alpha < \alphac$), sources and sinks with a particular wavenumber are selected. (iii) For periodic boundary conditions with $\alpha > \alphac$, we find travelling waves with perfect global polar order. (iv) For Neumann and Dirichlet boundary conditions, the phenomenology above $\alphac$ is different. For $\alpha$ slightly above $\alphac$, the system shows intermittent behaviour, with fluctuating domains of polar order. For larger $\alpha$, the system splits into subdomains with opposing polar order. (v) The transition point $\alphac \sim 0.60$ coincides with the crossover value $\alphax \sim 0.62$ predicted by the Eckhaus instability of the plane waves. This is in a stark contrast with 2D, where the transition point $\alphac \sim 0.28$ is much smaller than the Eckhaus threshold $\alphax \sim 0.58$.

The rest of the paper is organized as follows.
In \cref{sec:model}, we present our model and methods.
In \cref{sec:P-BC}, we present our results for periodic boundary conditions. We highlight the features of the source and sink solutions, wave number selection, defect density, polar order, and the disorder-to-order transition.
In \cref{sec:other-bc}, we consider Dirichlet and Neumann boundary conditions and discuss how the boundary conditions alter the phenomenology.
We conclude the paper with a discussion and future perspectives.
\section{Model and Methods\label{sec:model}}

We consider a minimal model of two conserved scalar fields $\phi_{1}(x,t)$ and $\phi_{2}(x,t)$ with non-reciprocal interactions on a one dimensional domain. The conserved dynamics of the complex scalar order parameter $\phi = \phi_1 + i \phi_2$ is governed by the continuity equation $\partial_t \phi = \partial_x^2\mu$, where $\mu = \delta \mathcal{F}/\delta \phi^{*} + i \alpha \phi$ is the non-equilibrium chemical potential. The equilibrium contribution to $\mu$ promotes phase-separation and is derived from the free energy functional $\mathcal{F} = \int \dd x \left(-|\phi|^{2}/2+|\phi|^{4}/4+|\nabla\phi|^{2}/2\right)$. The nonequilibrium part encodes the non-reciprocal interactions between the two species; quantified by the parameter $\alpha$. For $\alpha=0$, the model describes equilibrium phase-separation of two interacting species, and an initial disordered state coarsens towards a bulk-phase separated state. Any $\alpha\neq0$ sets the system out of equilibrium and it describes non-reciprocal phase-separation. A positive $\alpha$ implies that $\phi_1$ chases $\phi_2$, and the model is invariant under the simultaneous transformation $\alpha \to -\alpha$ and $\phi_{1,2} \to \phi_{2,1}$. Combining these terms, we obtain the following non-dimensional equation for the evolution of $\phi(x,t)$~\cite{rana2024a, rana2024b, saha2020, you2020}
\begin{align}\label{eq:nrch}
    \partial_t{\phi} &= \partial_x^2\left[(-1 + i \alpha)\phi + |\phi|^2\phi - \partial_x^2 \phi\right].
\end{align}

The homogeneous state $\phi=0$ is unstable to small perturbations and a linear stability yields that perturbations of the form $\delta\phi(k,t) \exp(ikx)$ evolve as
\begin{align}
    \partial_{t}{\delta\phi}(k,t) = k^2(1-i\alpha-k^2)\delta\phi(k,t), \label{eq:lin-stab}
\end{align}
which reveals that $\forall~|k| < 1$ the perturbations grow in an oscillatory manner. The behaviour of the system depends on $\alpha$, domain length $L$, and the choice of boundary conditions. We systematically study the phase-behaviour of \cref{eq:nrch} with varying strength of nonreciprocity $(\alpha)$ for various domain sizes $(L)$ and boundary conditions. We consider three different kinds of boundary conditions that are relevant to the phenomenon of phase separation~\cite{rodrigues1989}:
\begin{enumerate}[label={(\roman*)}]\label{tab:bc}
    \item Periodic boundary conditions (P-BC), where
        \begin{align}
            \phi(x+L) = \phi(x)~\mathrm{and}~\mu(x+L) = \mu(x).
        \end{align}
    \item Neumann boundary conditions (N-BC), where
        \begin{align}
            \partial_x\phi|_{0} = \partial_x\phi|_{L} = 0~\mathrm{and}~\partial_x\mu|_{0} = \partial_x\mu|_{L} = 0.
        \end{align}
    \item Dirichlet boundary conditions (D-BC), where we fix $\phi$ on boundaries and $\mu$ is flux-free as
        \begin{align}
            \phi(0) = \phi(L) = 0~\mathrm{and}~\partial_x\mu|_{0} = \partial_x\mu|_{L} = 0.
        \end{align}
\end{enumerate}
One can also consider fixed boundary conditions for $\mu$, but it is not compatible with the conservation of mass, thus we do not consider this case.

\begin{figure}[b]
    \centering{\includegraphics[width=\linewidth]{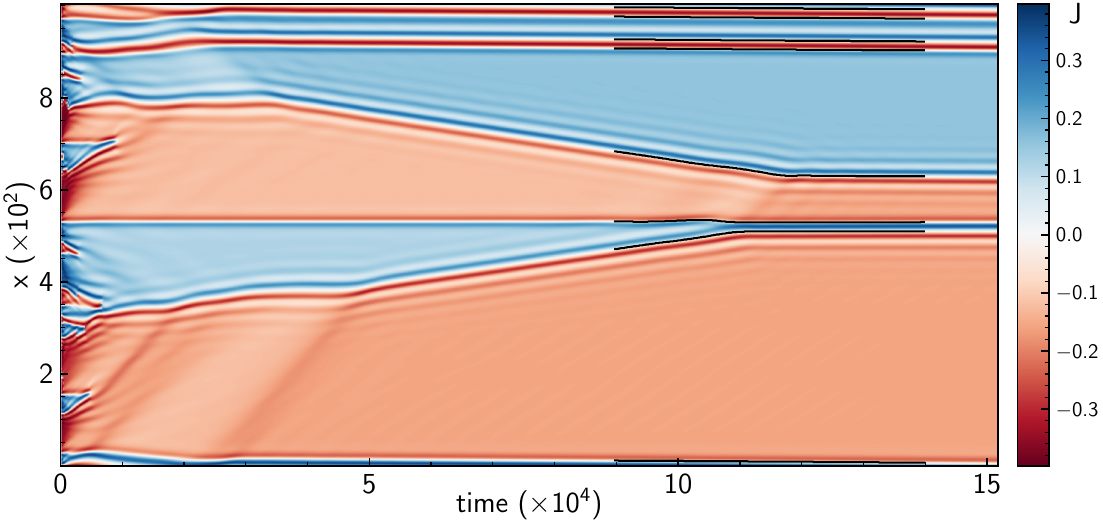}}
    \caption{\label{fig:Jkymograph} Kymograph plot of the polar order parameter $J(x,t)$ highlighting the evolution of disordered states. Defects are the zeros of $J(x,t)$, which appear white in the kymograph. We also mark the exact defect positions (small black points) at late times. Initially, numerous source-sink pairs are formed, which merge and the system settles into a stable configuration, which persists till the end of the simulations.
    }
\end{figure}

\begin{figure*}
    \centering{\includegraphics[width=\linewidth]{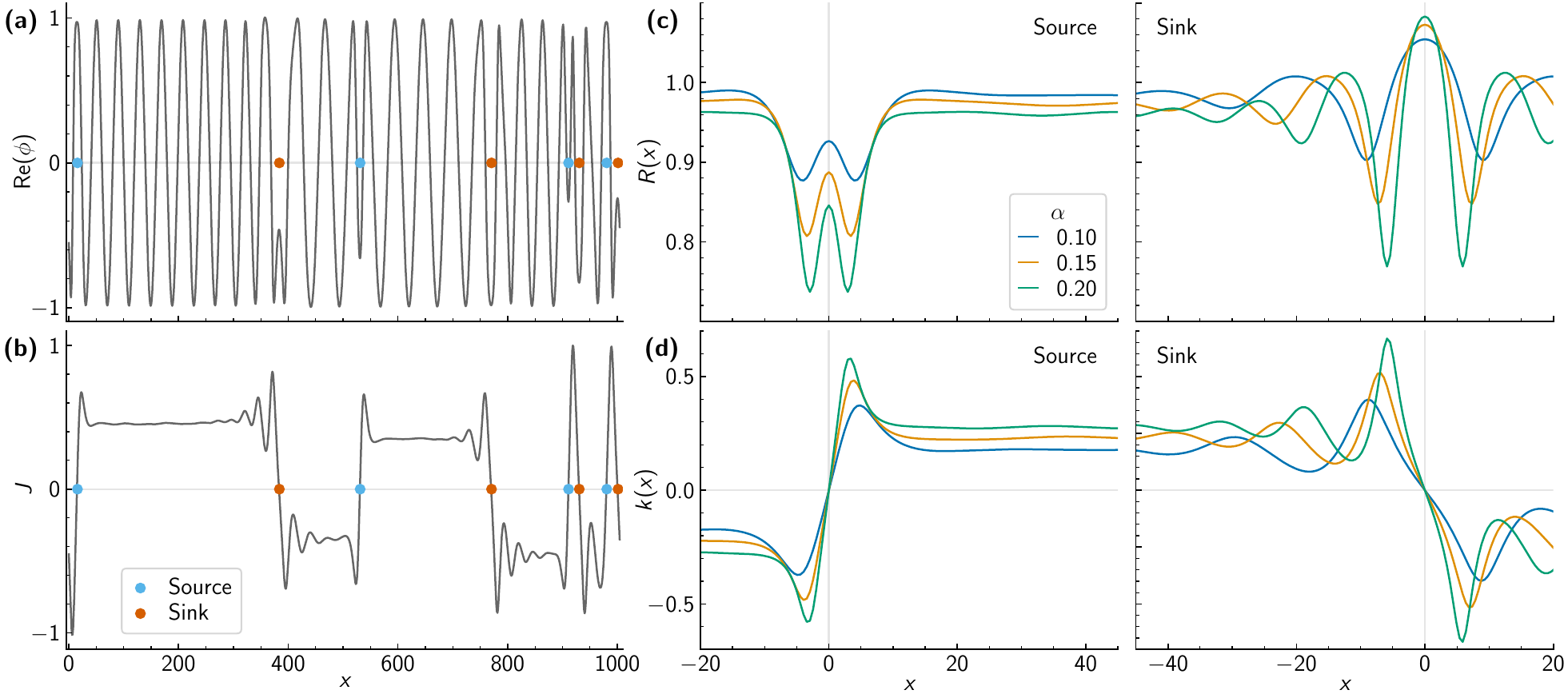}}
    \caption{\label{fig:defects}
        Sources and sinks in the 1D NRCH model.
        (a, b) Representative plot of $\real{\phi}$ and the polar order parameter $J$ (normalized by its maximum value), respectively.
        Sources (blue circles) and sinks (orange circles) are arranged in an alternating manner on the 1D domain.
        $J$ vanishes at defect centers and is constant far from defects, where the wavefront is that of a plane wave.
        (c, d) Plot of $R(x)$ and $k(x)$ for sources and sinks for different values of $\alpha$.
        For these plots, $x$ is measured from the center of the defects.
        At defect centers, $\partial_x R$ and $k(x)$ vanish.
        Far from the defects, $R(x)$ and $k(x)$ approach their plane wave forms.
}
\end{figure*}

To study the phenomenology, we numerically integrate \cref{eq:nrch} on a 1D domain of length $L$ that is discretized using $N$ points.
For P-BC, we use a pseudo-spectral algorithm with a second-order exponential time-differencing scheme~\cite{cox2002} for time marching.
For N-BC and D-BC, we perform the simulations using a Chebyshev polynomials based spectral Tau method implemented by the Dedalus framework~\cite{burns2020}.
Additionally, we use Dedalus to cross-check our results for the periodic domain. The source code to simulate the 1D NRCH model for all the boundary conditions considered here using Dedalus is available at \url{https://github.com/navdeeprana/DedalusScripts}.

We start all our simulations with a small perturbation to the homogeneous state.
i.e., $\phi(x,0)=\delta\phi(x)$, where at every point $x$ the initial perturbation $\delta\phi(x)$ is drawn from a uniform distribution.
Further, we impose  $\int_{0}^{L} \phi(x, 0) \dd x = 0$, which remains constant throughout the entire time evolution due to the conservation of mass.
Unless reported otherwise, we average over eight or more independent realizations of initial conditions for each data point reported in the paper.

\begin{figure}[b]
    \centering{\includegraphics[width=\linewidth]{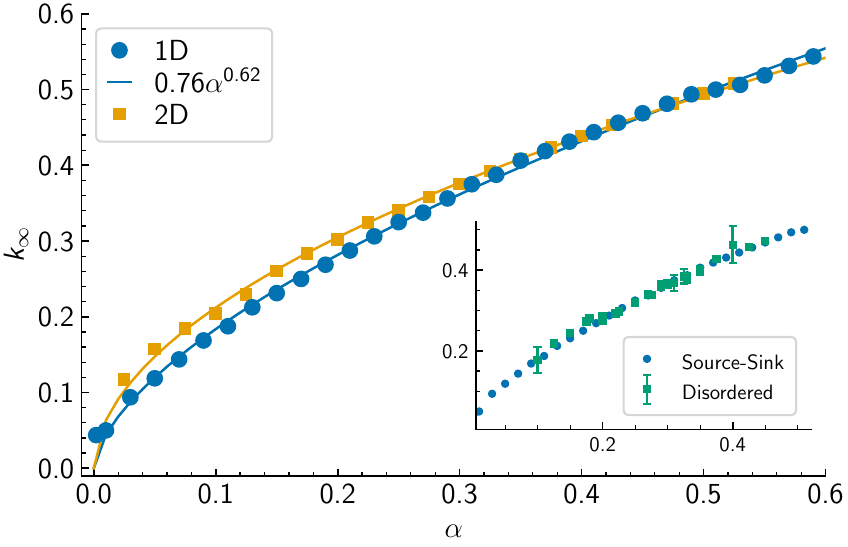}}
    \caption{\label{fig:wave_number} Wave number selection in 1D. For a given $\alpha$, sources and sinks with a particular wave number are selected. Using a least-square fit to power law forms, we obtain $\kl \sim \alpha^{0.6}$ (Solid black line). The selected wave number is comparable to the $\kl$ obtained for the targets in 2D (Orange circles)~\cite{rana2024a}. Inset: $\kl$ for a single source-sink pair is identical to the $\kl$ for a multi-defect disordered configuration.
    }
\end{figure}

\section{Periodic boundary conditions\label{sec:P-BC}}

For P-BC, the NRCH model \eqref{eq:nrch} admits travelling wave solutions of the form
\begin{align}\label{eq:wave}
    \phi(x,t) = R \, e^{i(k x - \omega t)},
\end{align}
where $k < 1$, $R^2 = 1-k^2$, and $\omega = \alpha k^2$.
For small $k$, travelling states are stable to small perturbations, however an Eckhaus instability kicks in at large $k$ and restricts the allowed wave number range to $0 \leq k^2 < 1/3$~\cite{aranson2002, saha2025, rana2024a, zimmermann1997}.
In our previous studies~\cite{rana2024a, rana2024b}, we have shown that in 2D, at low $\alpha$, disordered states evolves to defect configurations composed of spirals and targets.
In 1D, we find similar phenomenology.
A random state evolves into multi-defect configurations that persist up to the end of simulation time (see \cref{fig:Jkymograph} and \cref{fig:defects}).
We now discuss the properties of these states in detail.

\begin{figure*}
    \centering{\includegraphics[width=\linewidth]{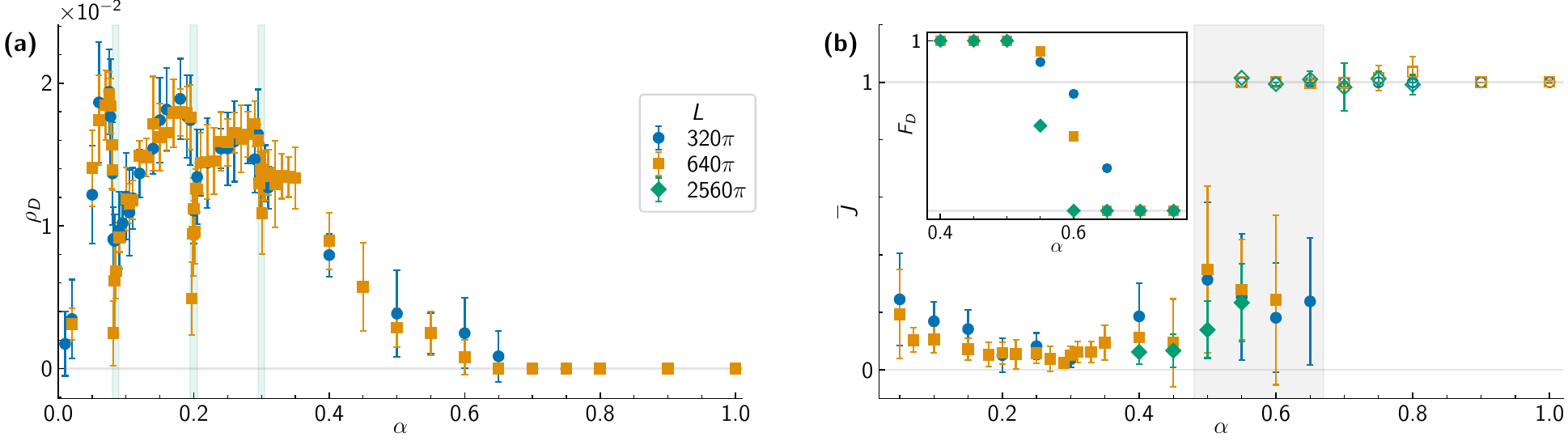}}
    \caption{\label{fig:P-BC_transition}
        (a) Defect density $\rho_D$ versus $\alpha$ for P-BC.
        Overall $\rho_{D}$ peaks around $\alpha \sim 0.2$ and vanishes for $\alpha \to 0$ and $\alpha \to \alphac$. Above $\alphac \sim 0.6$, $\rho_D=0$ which implies the travelling wave phase.
        At certain values of $\alpha \simeq 0.08, 0.2, 0.3$, which we call ``resonances" (shaded green), $\rho_{D}$ significantly dips.
        (b) Average polar order $\avj$ versus $\alpha$.
        For small $\alpha$, $\avj \sim 0$ and at large $\alpha$, $\avj \sim 1$.
        For $\alpha \sim 0$ or $\alpha \sim \alphac$ (grey shaded region), $\avj$ shows large fluctuations due to finite size effects. We compute $\bar{J}$ separately for defect-riddled states (solid markers) and travelling states (hollow markers). In the shaded region, some realizations show defects, some show travelling waves.
        Inset: The fraction of ensemble $(F_{D})$ that show defect states for $\alpha$ close to $\alphac$.
        Below $\alphac$, $F_{D} = 1$, above $\alphac$, all simulations transition to travelling waves, thus $F_{D} = 0$.
        Larger domains show sharper transitions.
        From these plots, we infer $\alphac \sim 0.6$.
    }
\end{figure*}

\subsection{Sources and Sinks\label{sec:defects}}
The 1D NRCH model \cref{eq:nrch} admits solutions of the form
\begin{align}\label{eq:defect}
    \phi(x,t) = R(x) \, e^{i\left(Z(x) - \omega t \right)},
\end{align}
where $x$ is measured from defect location, $R(x)$ is the amplitude, $Z(x)$ is the phase, and $k(x) \equiv \dd Z/\dd x$.
They are the sources and sinks of travelling waves and we shall collectively call them defects.
In \cref{fig:defects}(a), we plot a representative multi-defect configuration obtained from our P-BC simulations.
The sources and sinks are arranged in an alternating manner on the 1D domain and the number of sources is always equal to the number of sinks.
It is easy to identify the defect locations from the polar order parameter
\begin{align}\label{eq:polarorder}
    J(x,t) =\mathrm{Im}(\conj{\phi}\partial_x\phi)  =  \frac{1}{2 i}\left(\conj{\phi} \partial_x \phi - \phi\partial_x\conj{\phi}\right).
\end{align}
As shown in \cref{fig:defects}(b), at a source (sink), $J$ vanishes with a positive (negative) slope, thus at a given time $t$, the number of defects is equal to the number of zeros of $J(x,t)$.
Quantities similar to $J(x,t)$ have been used earlier to analyse the defect dynamics in systems with complex order parameter~\cite{angheluta2012}.
Substituting the defect ansatz in the definition of $J(x,t)$, we further find that at the defect core $\partial_x (R^2) = \partial_x(\phi \conj{\phi}) = 2 \conj{\phi}\partial_x\phi = 2 \phi \partial_x \conj{\phi}$.
Thus, at the defect positions, %
\begin{align}
    \partial_x R(x)\Big|_{x=0} = 0,~\mathrm{and}~k(0) = 0.
\end{align}
In \cref{fig:defects}(c,d), we plot the profiles of $R(x)$ and $k(x)$ for sources and sinks for different $\alpha$.
For a source, $k(x) > 0 (< 0)$ when $x > 0 (< 0)$, and vice versa for a sink.
It is evident that $R(-x) = R(x)$ and $k(-x) = - k(x)$, which implies that for small $x$, $R(x) \sim a_0 - a_2 x^2$ and $k(x) \sim b_1 x - b_3 x^3$.
Far away from the defect center, the wavefront approaches that of a travelling wave meaning for $x \gg 1$, $k(x) \to \pm\kl$ for a source and $k(x) \to \mp \kl$ for a sink (see \cref{fig:defects}(a)).
The amplitude $R(x) \to \sqrt{1-\kl^2}$ also becomes constant and to ensure proper oscillations, we require $\omega = \alpha \kl^2$.
Note that the sources are the direct 1D analogue of targets~\cite{rana2024a, rana2024b} and topologically charged defects do not exist in 1D.

\begin{figure*}
    \centering{\includegraphics[width=\linewidth]{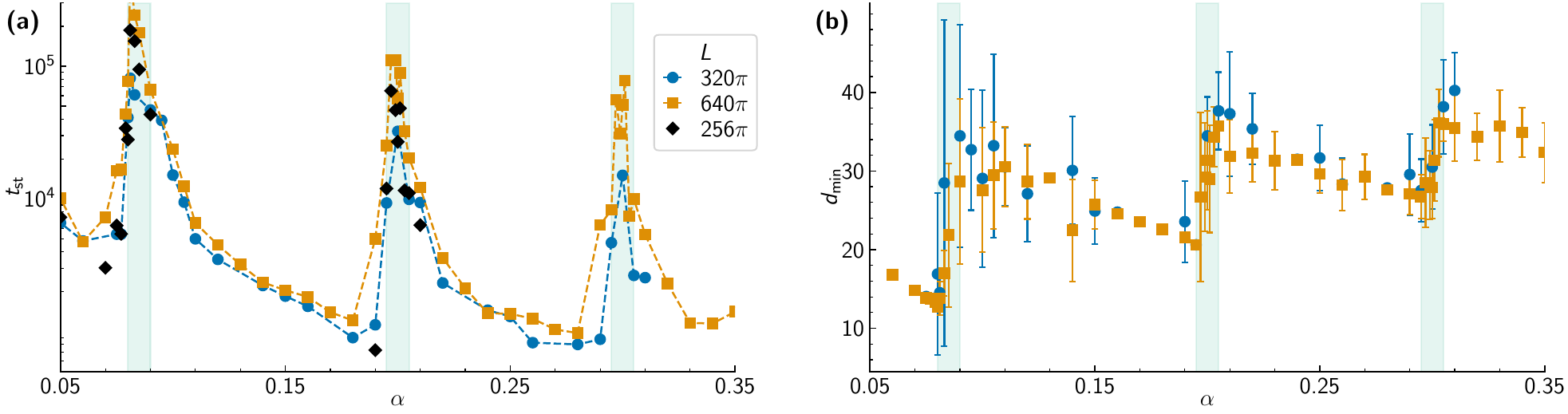}}
    \caption{\label{fig:P-BC_resonances}
        Analysis of resonances.
        (a) Time ($t_{\text{st}}$) after which the system reaches the steady-state value of $\rho_{D}$.
        At resonances, defect merger events continue for long times, thus $t_{\text{st}}$ is orders of magnitude larger. To verify that the resonances are not the effect of domain size, we also plot results for $L=256\pi$ (black points), which shows similar behaviour.
        (b) Minimum inter-defect separation $\dmin$ versus $\alpha$.
        As defects keep merging for long times, we find sharp jumps in $\dmin$ in the resonance regions.
    }
\end{figure*}

\subsection{Wave number selection}

As we show in \cref{fig:wave_number}, for a given $\alpha$, sources and sinks with a particular value of the asymptotic wave number $\kl$ are selected.
$\kl$ increases monotonically with $\alpha$, and from a least-square fit, we obtain $\kl \sim \alpha^{0.6}$.
Furthermore, $\kl$ is comparable to the wave number selected for targets in 2D.
In \cref{fig:wave_number}(inset), we verify that $\kl$ for a single source-sink pair configuration is the same as the $\kl$ obtained from multi-defect configurations that evolved from disordered initial conditions.
A monotonically increasing $\kl$ predicts a crossover value of non reciprocity $\alphax$, such that when $\alpha \to \alphax$, $\kl \to 1$, and $\Rl \to 0$.
This implies that the defect solutions will cease to exist for $\alpha \geq \alphax$.
Using a least-square fit, we find $\alphax \sim 1.5$.
However this prediction of $\alphax$ is not entirely correct as it does not account for the Eckhaus instability.
Since the plane waves generated by the defects are unstable for $k^2 \ge 1/3$~\cite{aranson2002, saha2025, rana2024a, zimmermann1997}, we expect that the defect solutions will vanish for $\kl^2 \ge 1/3$ which yields $\alphax \sim 0.62$. Thus wavenumber selection and the Eckhaus instability restrict the defect solutions to the range $0 < \alpha < 0.62$. It is interesting to note that this threshold for the stability of the defects is directly connected to the onset of the emergence of global polar order (see below). 

\subsection{Defect density and the Global polar order}

We now focus on the statistical properties of the multi-defect configurations.
Using the fact that $J=0$ at a source or a sink, we define the 1D defect density $\rho_D$ simply as the number of zeros of $J$ per unit length.
In \cref{fig:P-BC_transition}(a), we plot $\rho_D$ in the steady state for different values of $\alpha$.
Since $\alpha=0$ is the equilibrium limit of the model, $\rho_D$ vanishes as $\alpha \to 0$.
Furthermore, $\rho_D$ vanishes above $\alphac \sim 0.6$ and instead we find travelling waves.
This marks the onset of disorder-to-order transition that was also observed in 2D~\cite{rana2024a}.
To better analyse the transition, in \cref{fig:P-BC_transition}(b), we plot the average polar order $\avj$ versus $\alpha$, in the steady state.
$\avj$ is defined as
\begin{align}\label{eq:average_polar_order}
    \avj = \frac{1}{(\qp - \qp^3)}\left|\left<J(x,t)\right>\right|,
\end{align}
where $\left<\ldots\right>$ implies averaging over space and time in the steady-state, and we use the dominant wave number of the patterns $\qp$ for normalization.
For travelling waves, $J = R^2 k = k - k^3$, and $\qp = k$, which sets $\bar{J} = 1$ for all travelling waves, irrespective of their wave number.
For defects, $\qp=\kl$ is a suitable choice.
Since defects are sources or sinks of travelling waves in both the directions, we expect $\avj$ to vanish in the presence of defects.
For $\alpha < \alphac$, $\avj$ is close to zero and for $\alpha > \alphac$, consistent with the travelling states, $\avj \sim 1$.

For very small $\alpha$, or for $\alpha \lesssim \alphac$, $\rho_{D}$ is small, which implies that there are only a handful of defects in the domain. This can lead to finite size effects and large fluctuations that are reflected in the plot of $\avj$. However, these effects become smaller as the domain size increases.

Another consequence of the finite domains is that in the vicinity of $\alphac$ (grey shaded region), the steady state varies across the independent realizations of the initial conditions.
Some of the simulations transition to travelling waves while others show defects.
Thus, to find a better estimate of $\alphac$, we compute the fraction of simulations $(F_{D})$ that show defect states (see \cref{fig:P-BC_transition}(b,inset)).
For $\alpha \lesssim 0.4$, $F_{D} = 1$ which implies that the entire ensemble shows defect states.
For $\alpha > 0.6$, all the simulations for the largest domains end up in travelling wave states.
From the plots of $\rho_D$ and $\avj$, we conclude that the disorder-to-order transition in 1D occurs at $\alphac \sim 0.6$, which is in agreement with the crossover threshold $\alphax \sim 0.62$ predicted by the wave number selection and the Eckhaus instability. 

To understand this result, we note that the travelling wave solutions of NRCH are linearly stable at all values of $\alpha$ \cite{saha2025}, and therefore, the only reason they might not be adopted at relatively smaller values of $\alpha$ is the emergence of relatively stable defects that form an arrested defect network \cite{rana2024a,rana2024b}. Once these defects are destabilized (beyond $\alphax \sim 0.62$) due to Eckhaus instability, the only stable solutions that remain will be the travelling waves. While the above argument, which supports $\alphac \simeq \alphax$, provides a simple and comprehensive explanation for the 1D phase diagram, we note that this scenario does not hold for the disorder-to-order transition in 2D, where $\alphac \sim 0.28$ is much smaller than $\alphax \sim 0.58$~\cite{rana2024a}. This suggests that in 2D there must be other effects (not present in 1D) that can destabilize the defects before the Eckhaus instability occurs. These effects could originate from the role of geometry of the domains, the existence of different types of defects, etc, and highlight the role of dimensionality in the phenomenology of the NRCH model.

\subsection{Resonances}

A careful analysis of the $\rho_{D}$~versus~$\alpha$ plot reveals nontrivial features of the defect configurations.
We find that $\rho_D$ shows sharp minima at certain values of $\alpha$ which we call ``resonances" (see \cref{fig:P-BC_transition}(a)).
A visual inspection of the evolution of the defect configurations reveals that at resonances, the defects keep merging even at longer times resulting in smaller defect density.
For further analysis, we compute the time it takes for the defect configurations to reach a steady state ($t_{\mathrm{st}}$), where $t_{\mathrm{st}}$ is defined as the time after which no defect merger event takes place and the defect density becomes constant.
We find that for the majority of $\alpha$-space, defects attain a steady state relatively quickly, however in the resonance regions, $t_{\mathrm{st}}$ can be order of magnitude higher (see \cref{fig:P-BC_resonances}(a)).
Consistent with these observations, we also observe sharp rise in the nearest neighbor separation $\dmin$ for the resonances (see \cref{fig:P-BC_resonances}(b)).
Similar phenomenon, where $d_{\mathrm{min}}$ rises sharply and decreases gradually was also observed in the study of pairwise defect interactions in 2D~\cite{rana2024b}.

\section{Effect of Boundary Conditions\label{sec:other-bc}}

We now consider the effects of varying boundary conditions on the phenomenology of the NRCH model.
As discussed in \cref{sec:model}, in addition to the periodic boundary conditions, we consider Neumann and Dirichlet boundary conditions.
Traveling waves of the form \eqref{eq:wave} are incompatible with both N-BC and D-BC, thus we do not expect states with complete polar order to emerge beyond $\alphac$.
Further, for both N-BC and D-BC, polar order vanishes at the boundaries, i.e.,
\begin{align}
    J(x=0,t) = J(x=L,t) = 0~\forall~t,
\end{align}
as either $\phi$ or $\partial_{x}\phi$ vanish at the boundaries.
Thus, the boundaries act as additional sources or sinks and in our analysis, we ignore the defects on the boundaries and only consider defects that are of the form described in \cref{fig:defects} and are located inside the domain within $x \in (x_{\mathrm{pad}}, L - x_{\mathrm{pad}})$.
For both N-BC and D-BC, we heuristically choose $x_{\mathrm{pad}} = 50$, which we find, filters the boundary defects very well.
\begin{figure}
    \centering{\includegraphics[width=\linewidth]{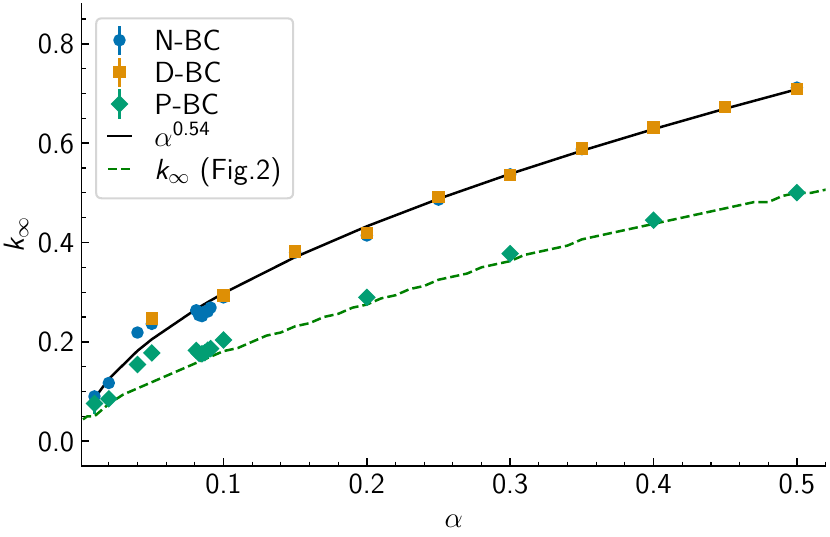}}
    \caption{\label{fig:wavenumberBC}
        Comparison of $\kl$ for different $\alpha$ for $\alpha < \alpha_{c}$ for different boundary conditions.
        For a given $\alpha$, $\kl$ is higher for both N-BC and D-BC when compared to the $\kl$ for the P-BC.
        As a check, we also plot the $\kl$ obtained from Dedalus for the P-BC, which matches with our own simulations (dashed black line).
    }
\end{figure}

\begin{figure*}
    \centering{\includegraphics[width=\linewidth]{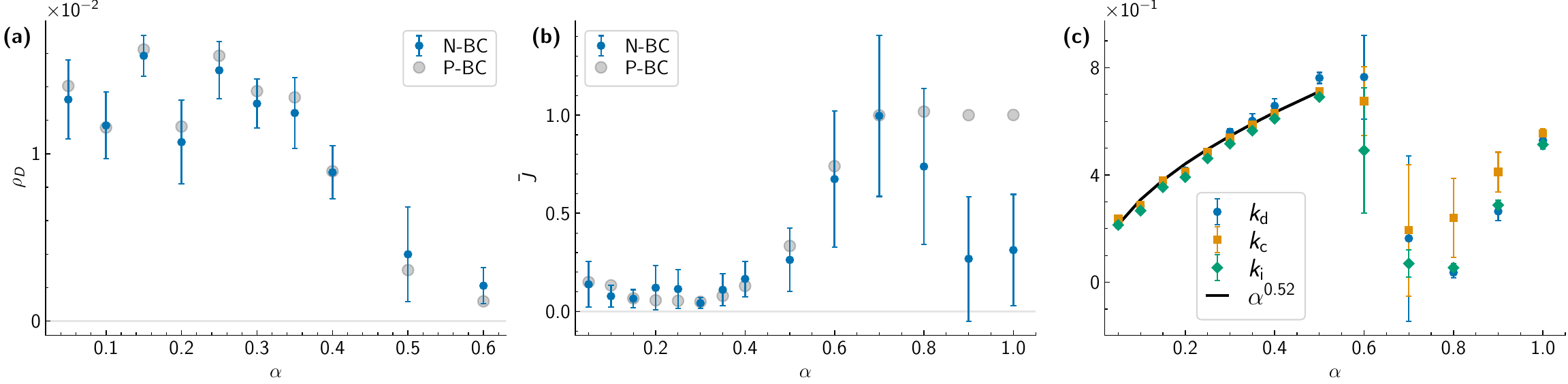}}
    \centering{\includegraphics[width=\linewidth]{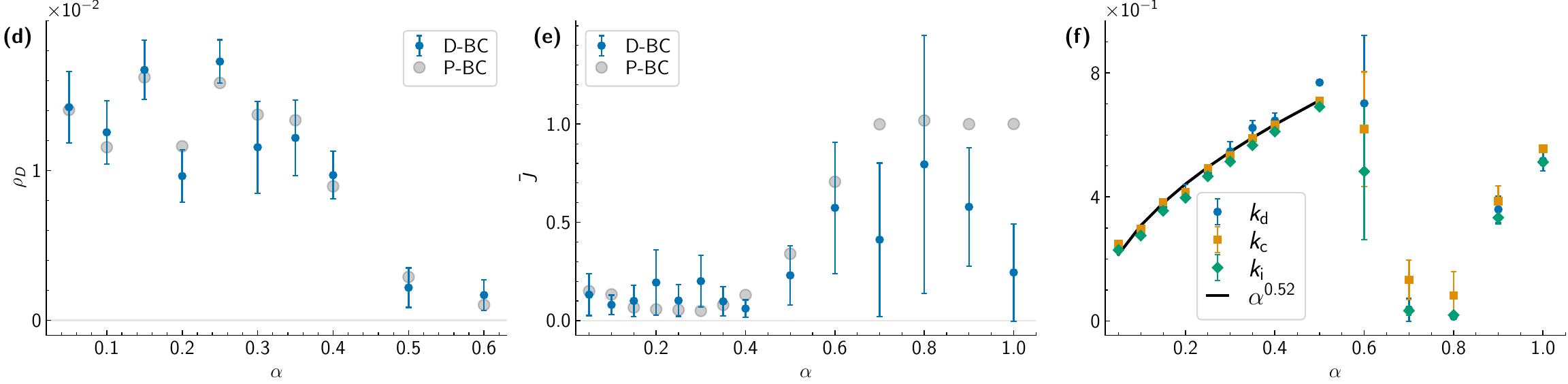}}
    \caption{\label{fig:BC_phenomenology}
        Phenomenology for the N-BC (a-c) and D-BC (d-f).
        For $\alpha < \alphac$, the phenomenology is similar to the P-BC.
        (a, d) Plot of average defect density $\rho_{D}$ versus $\alpha$, which is consistent with the statistics for P-BC within the error bars.
        (b, e) Plot of the average polar order.
        For $\alpha < \alphac$, we find $\avj \sim 0$.
        For $\alpha > \alphac$, the system shows partial order.
        Near the transition point, we observe large intermittent fluctuations, which are reflected in the large fluctuations in $\avj$ as well.
        (c, f) Comparison of different length scales.
        For $\alpha < \alphac$, $\kl$ is the dominant length scale and all other scales converge to it.
        Near the transition, we find large disagreements as well as large fluctuations and different length scales, which once again converge at larger $\alpha \sim 0.9$.
    }
\end{figure*}

\subsection{Defects below $\alphac$}
Below $\alphac$, we once again find defect configurations. As shown in \cref{fig:wavenumberBC}, similar to the P-BC, for both N-BC and D-BC, defects with a particular wave number are selected. However, wave number selection is affected by the presence of boundaries; for a given $\alpha$, the selected wave number is higher than that for the P-BC and a least-square fit yields $\kl = \alpha^{0.52}$. Thus, one expects that defect solutions will vanish around $\alpha \sim 1$. It is tempting to use the Eckhaus instability to arrive at a crossover threshold of nonreciprocity; $\alphax \sim 0.33$ for N-BC and D-BC. However, we note that the instability analysis assumes travelling wave states, which are not permitted solutions for these boundary conditions. Our numerical analysis suggests that the defect states persist up to $\alphac \sim 0.6$.

The phenomenology below $\alphac$ is qualitatively similar to the P-BC.
The defect density $\rho_D$ agrees with the results from the P-BC (see \cref{fig:BC_phenomenology}(a, d)).
Further, the average polar order vanishes for defect states.

\subsection{Fluctuating polar order beyond $\alphac$}

\begin{figure*}
    \centering{\includegraphics[width=0.9\linewidth]{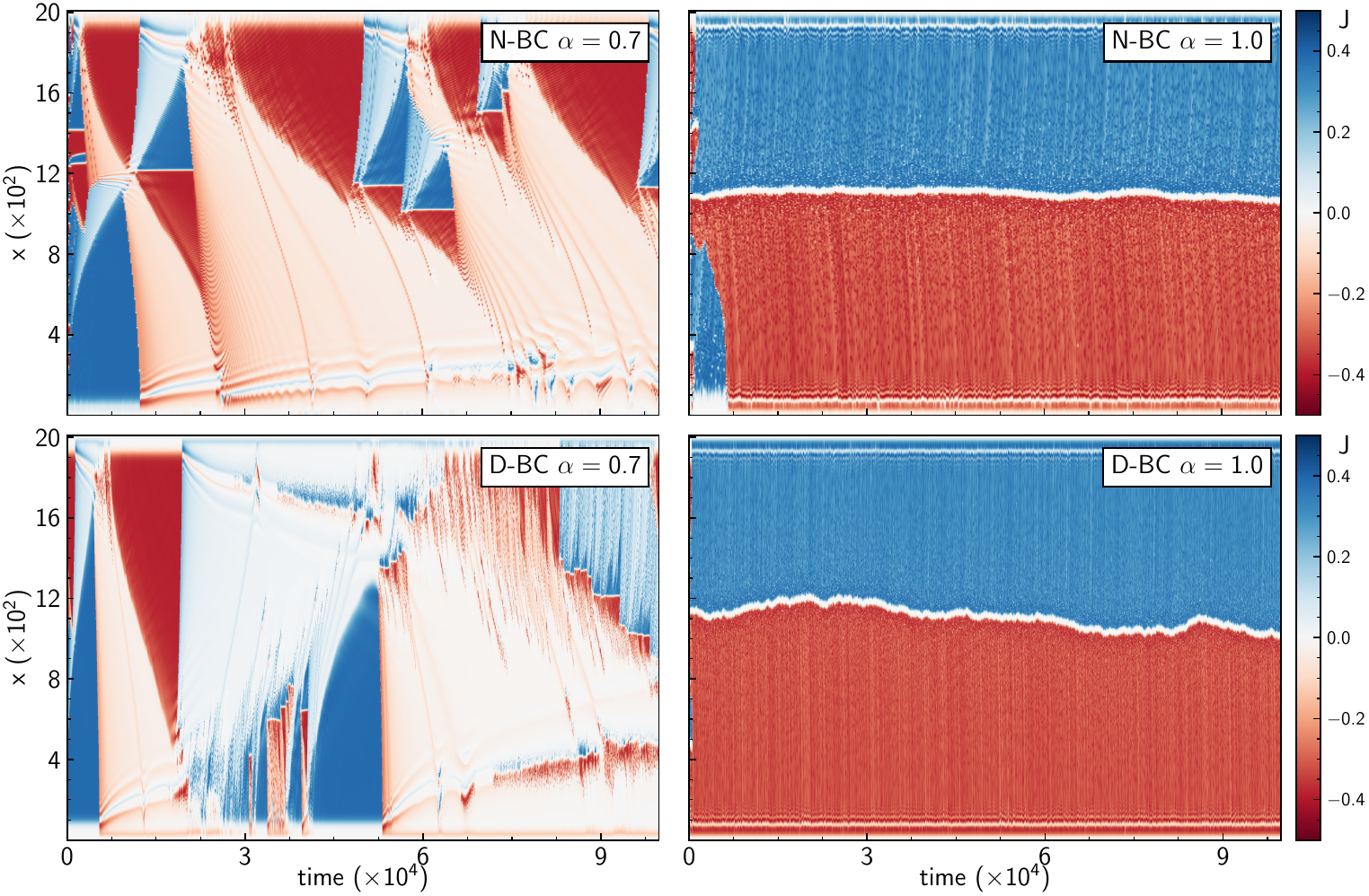}}
    \caption{\label{fig:polar_order_bc} Kymographs of $J(x,t)$ for N-BC and D-BC for $\alpha = 0.70$ (slightly above $\alphac$) and for $\alpha=1.0$.
        Closer to the transition point, we observe large transient and intermittent patches of polar order in the domain that keep emerging and vanishing.
        For large $\alpha$, both N-BC and D-BC show that the system develops two partitions with polar order in opposite directions.
        The position of the partition fluctuates with time.
    }
\end{figure*}

Since travelling waves are only permitted for P-BC, we expect the phenomenology for $\alpha > \alphac$ to vary significantly for N-BC and D-BC. Indeed, beyond $\alphac$, the system does not attain a single travelling wave phase. For $\alpha \gtrsim \alphac$, both N-BC and D-BC show intermittent behaviour. The emergence of global polar order is thwarted by the incompatibility of travelling bands with the boundary conditions. Thus, the system does not settle down, but shows transient, fluctuating states with patchy polar order that sustain over long times (see \cref{fig:polar_order_bc}).

In the case of P-BC, a single wavenumber dominates the steady states, for defect configurations below $\alphac$ it is the selected wavenumber $\kl$, for the travelling states above $\alphac$, it is the wave number of the travelling wave. For N-BC and D-BC, we find that the intermittent patchy ordered states exhibit fluctuations at multiple wave numbers. To verify this, we compute various relevant scales using the structure factor $S(k,t)$:
\begin{enumerate}[label={(\roman*)}]\label{tab:bc}
    \item The dominant mode $\kd$ defined as the wavenumber at which $S(k)$ peaks, i.e, $\kd= \mathrm{argmax}_{k} S(k)$.
    \item The coarsening mode $\kc= \sum_{k} k S(k) / \sum_{k} S(k)$
    \item The integral mode $\ki = \sum_{k} S(k) /\left(\sum_{k} S(k) / k\right)$.
\end{enumerate}

Here, $S(k, t) = |\psi(k)|^{2}$ is the structure factor, where $\psi(k,t) = \sum_{x}\psi(x,t)\exp(-ikx)$ is the Fourier transform of the windowed field $\psi(x,t) = w(x) \phi(x, t)$. Note that we apply the Hanning window $w(x) = \mathrm{sin}^{2}(\pi x / L)$ on the data for the N-BC and D-BC to avoid spurious amplitude errors arising from the aperiodicity.

As shown in \cref{fig:BC_phenomenology}(c,f), $\kd$, $\kc$, and $\ki$ are in excellent agreement with each other and match with $\kl$ for $\alpha<\alphac$, emphasizing that $\kl$ governs the dynamics for defect configurations. Slightly above $\alphac$, different modes disagree with each other, and exhibit large fluctuations, signifying that the transient states show fluctuations at multiple length scales. Consequently, it is not possible to choose a suitable normalization for the average polar order; i.e., a suitable choice for $q_{p}$ in \cref{eq:average_polar_order} is not available for these states, which is also reflected in large fluctuations of $\avj$ (see \cref{fig:BC_phenomenology}(b,d)) where we have used $q_{p} = \kc$.

For $\alpha > \alphac$, the system solves the incompatibility with the travelling waves by creating two domains with waves travelling in the opposite direction (see \cref{fig:polar_order_bc}). These domains are separated by a partition where $|\phi|$ vanishes.
and in addition to the domain boundaries, it acts as a source or sink for the travelling waves.
The partition can form anywhere inside the domain and fluctuates slowly with time.
The distribution function of $J(x,t)$ is primarily bimodal, where the strength of the two peaks depends on the location of partition in the domain.
As a consequence, we find large non zero polar order when the partition is closer to one of the domains, or small polar order if it is close to the middle of the simulation domain.
Finally, various scales, $\kd$, $\kc$, and $\ki$, all converge to similar values.

\section{Conclusions}

We have presented a systematic study of the disorder-to-order transition for the 1D NRCH model.
In 1D, defect solutions are the sources and sinks of travelling waves.
Sources are analogous to the two-dimensional topologically neutral targets, whereas sinks are similar to the disclination lines where the waves emanated from sources meet.
For a given $\alpha$, defects select a unique wave number $\kl$ that increases monotonically with $\alpha$.
Wave number selection predicts a crossover threshold $\alphax \sim 0.62$ above which the defect solutions cannot exist.

At small $\alpha$, disordered initial states evolve into multi-defect configurations, where sources and sinks are arranged in an alternating manner.
With increasing $\alpha$, overall the defect density increases, peaks at $\alpha \sim 0.2$, and then vanishes at $\alphac \sim 0.6$, which marks the onset of disorder-to-order transition.
In accordance with this observation, we find that there is no significant average polar order for $\alpha < \alphac$, but travelling states for $\alpha > \alphac$ show perfect order in the long time steady state.
In a direct contrast to 2D, where the transition occurs at $\alphac \sim 0.28 \ll \alphax \sim 0.58$, the transition point in 1D, $\alphac \sim 0.6$ is in close agreement with the crossover threshold $\alphax \sim 0.62$ predicted by wavenumber selection.
A closer inspection of the defect dynamics reveals that at certain resonance values of nonreciprocity below $\alphac$, $\rho_D$ shows sharp minima, which is a consequence of the fact that for these values of $\alpha$, defect merger events continue to occur even at very long times.
Additionally, the average inter-defect separation rises sharply for these $\alpha$ values, which was also observed in the study of pairwise defect interactions in 2D~\cite{rana2024b}.

Our numerical simulations with N-BC and D-BC, which are inconsistent with the travelling waves, show a different flavor of the disorder-order transition.
The defect dynamics at $\alpha < \alphac$ is similar to the P-BC, but the boundary conditions significantly affects the phenomenology above the transition point.
Slightly above the threshold, we find intermittent patches of polar order.
For larger values of $\alpha$, the system partitions into two subdomains which sustain over time.

To conclude, our study shows that for the P-BC NRCH model shows similar phenomenological behaviour in both 1D and 2D, with a crucial difference that the disorder-to-order transition occurs at the crossover value predicted by wave number selection in 1D.
We further show that boundary conditions can significantly affect the phenomenology.
These findings call for future studies to further investigate the defect dynamics and defect interactions for conserved systems with non-reciprocal interactions.
In particular, it will be interesting to explore what determines the value of $\alphac$ in 1D and 2D and to find out the mechanism that drives the resonances for certain values of $\alpha$.

\bibliography{Bibliography,biblio}
\bibliographystyle{apsrev4-2}

\end{document}

%% file: abbreviations.tex
\newcommand{\dd}{\mathop{}\!\mathrm{d}}

\newcommand{\real}[1]{\text{Re}\left(#1\right)}

\newcommand{\conj}[1]{\bar{#1}}

\newcommand{\Rl}{R_{\infty}}
\newcommand{\kl}{k_{\infty}}
\newcommand{\qp}{q_{p}}
\newcommand{\alphac}{\alpha_{c}}
\newcommand{\alphax}{\alpha_{\times}}
\newcommand{\avj}{\bar{J}}
\newcommand{\dmin}{d_{\mathrm{min}}}
\newcommand{\kd}{k_\mathrm{d}}
\newcommand{\kc}{k_\mathrm{c}}
\newcommand{\ki}{k_\mathrm{i}}